\documentclass[runningheads]{llncs}
\usepackage[inline]{enumitem}
\usepackage{graphicx,microtype,adjustbox,subfiles,multirow}

\newif\ifdraft\draftfalse
\drafttrue  
\PassOptionsToPackage{
  commentmarkup=todo,
  todonotes={textsize=tiny,
      textwidth=2.5cm
    },
  commandnameprefix=always}{changes}
\ifdraft
\else
  \PassOptionsToPackage{final}{changes}
\fi
\usepackage{changes}
\usepackage[backref=page]{hyperref}

\usepackage{amsmath}
\usepackage{amssymb}
\usepackage{amsfonts}

\def\real{\mathbb R}

\DeclareMathOperator{\argmin}{argmin}

\def\comment#1{}

\def\eqref#1{(\ref{#1})}
\def\Beq#1\Eeq{
  \begin{equation}
    #1
  \end{equation}
}
\def\Beqo#1\Eeqo{
  \begin{equation*}
    #1
  \end{equation*}
}
\def\Beqs#1\Eeqs{
  \begin{align}
    #1
  \end{align}
}
\def\Beqso#1\Eeqso{
  \begin{align*}
    #1
  \end{align*}
}

\PassOptionsToPackage{dvipsnames}{xcolor}
\usepackage{xcolor}
\usepackage{booktabs}

\definechangesauthor[name=Paweł, color=red]{PW}
\definechangesauthor[name=Radek, color=blue]{RN}
\definechangesauthor[name=Adam, color=green]{AM}
\definechangesauthor[name=Piotr, color=orange]{PS}
\definechangesauthor[name=Daniel, color=yellow]{DC}

\definechangesauthor[name=TODO, color=magenta]{TODO}
\definechangesauthor[name=Question, color=magenta]{Q}

\graphicspath{ {./img_} }

\begin{document}
  \title{Graph Vertex Embeddings: Distance, Regularization and Community Detection}
  %
  %
  \author{Radosław Nowak\inst{1} \and
    Adam Małkowski \and
    Daniel Cieślak\inst{1} \and
    Piotr Sokół\inst{1} \and
    Paweł Wawrzyński\inst{1}
    }
  \authorrunning{R.
    Nowak et al.
  }
  %
  \institute{${}^1$IDEAS NCBR, Chmielna 69, 00-801 Warsaw, Poland,
    \url{https://ideas-ncbr.pl} }
  %
  \titlerunning{Graph Vertex Embeddings}
  \maketitle
  \begin{abstract}
    Graph embeddings have emerged as a powerful tool for representing complex network structures in a low-dimensional space, enabling \comment{the use of} efficient methods that employ the metric structure in the embedding space as a proxy for the topological structure of the data.
    In this paper, we explore several aspects that affect the quality of a vertex embedding of graph-structured data.
    To this effect, we first present a family of flexible distance functions that faithfully capture the topological distance between different vertices.
    Secondly, we analyze vertex embeddings as resulting from a~fitted transformation of the distance matrix\comment{,} rather than as a~direct result of optimization.
    Finally, we evaluate the effectiveness of our proposed embedding constructions by performing community detection on a host of benchmark datasets.
    The reported results are competitive with classical algorithms that operate on the entire graph while benefitting from a substantially reduced computational complexity due to the reduced dimensionality of the representations.
    %
    %
    %
    \keywords{Graphs  \and Embeddings \and Graph drawing \and Community detection.}
  \end{abstract}
  \section{Introduction}
   
    Low-dimensional metric embeddings of non-metric data play a crucial role in various domains of computer science, e.g.:
   \begin{enumerate*}[label=\textbf{\alph*})]
   	\item machine learning, where probabilistic generative models are constructed to capture variability through low dimensional factors~\cite{vincent2010,kipf2016};
   	\item natural language processing, where symbolic/text data is represented vectorially in order to facilitate learning of statistical dependencies~\cite{vaswani2017};
   	\item information retrieval, where embeddings allow for efficient search~\cite{beygelzimer2006};
   	\item data visualization, where complex, high-dimensional data is represented in a two- or three-dimensional space~\cite{bohm2022}.
   \end{enumerate*}
   Crucially, low-dimensional embeddings of data mitigate the computational and statistical challenges collectively referred to as the \emph{curse of dimensionality}.
   
   Graph embeddings present a particularly interesting potential application since the data is inherently non-metric and high-dimensional while simultaneously being of tremendous practical interest, due to the ubiquity of graphs in various domains, e.g.: social networks, biological networks, energy grids, and knowledge graphs~\cite{jumper2021,morselligysi2021}.
   Addressing the challenges posed by `graph problems' requires faithful and compact representations of the data, i.e. they should retain information about important structural properties and allow flexible, computationally efficient use for downstream processing/tasks.
   \\
   
   Graph embeddings have been widely studied; past approaches include spectral embeddings, graph kernels, multi-dimensional scaling (MDS), locally linear embedding (LLE), and Laplacian eigenmaps (LE)\cite{wu2022c,agrawal2021}.
   Recently, graph neural networks (GNNs) have emerged as an empirically successful model class, which offers improved scalability and more flexible processing of the embeddings~\cite{wu2022c}.
   \\

   In this work, we introduce a novel method for representing graph-structured data in low-dimensional metric spaces, which combines the efficiency of optimizati\-on-based embeddings and the expressiveness of neural network-based approximate to accurately reflect the topological distances within the graph.
   By framing the embedding of vertices as an optimization problem, we parametrize the embedding using a small neural network, to regularize the resulting representation.
   Furthermore, our formulation can accommodate various distance functions, which allows us to adapt the geometry of the embedding space to better reflect the structure of the original graph.
   The resulting embeddings offer a compact yet faithful representation, which combined with off-the-shelf clustering algorithms allow us to effectively identify communities within the graph.
   Our approach not only ensures a more favorable computational and statistical scaling, comparable to that of Graph Neural Networks (GNNs) but also provides a highly expressive representation of the graph structure, capturing the intricate relationships and distances within the data.
   
  \section{Related work}
   
   The body of research on embeddings is extensive and has a long history.
   Consequently, we begin by recalling foundational results before providing a concise overview of the methodologies presently in use.
   For a comprehensive review see~\cite{wu2022c,zhang2021c}.
   
   Historically, the graph embedding problem has been initially studied in the context of dimensionality reduction and data visualization, while preserving important structural properties of the data.
   Seminal examples of such methods include PCA, graph kernels \cite{borgwardt2020}, Laplacian eigenmaps \cite{Belkin2001}, {Isomap} \cite{Tenenbaum2000}, LLE \cite{Roweis2000}, maximum variance unfolding \cite{Weinberger2006}, t-SNE \cite{maaten2008}, LargeVis \cite{Tang2016}, UMAP \cite{McInnes2020}, and the latent variable models (LVM).
   
   \cite{agrawal2021} introduced \emph{minimum distortion embeddings} as a unifying framework that subsumes all of the previously mentioned methods, except t-SNE and LVM.
   In their formulation, low-dimensional embeddings are constructed by minimizing the distortion of pairwise distances between data points in the original and the embedded spaces.
   Independently, \cite{Johnson1984,Bourgain1985,linial1995,indyk2001} have shown that an \(m\)-dimensional embedding of graph with \(n\) vertices incurs a distortion of order \(\mathcal{O}\left(\log n\right)\), where \(m\) is \(\mathcal{O}\left(\log^{2} n\right)\).
   %
   %
   Remarkably, the landmark results of Johnson, Lindenstrauss, and Bourgain~\cite{linial1995} not only provide the mathematical foundation of low-dimensional representations of data but also have led to the development of fast, randomized algorithms.
   \\
   
   Currently prevalent algorithmic approaches present a variety of design options.
   An initial consideration requires the specification of a geometry for the embedding space; potential choices include hyperbolic~\cite{nickel2017,klimovskaia2020,sala2018}, spherical~\cite{agrawal2021}, and vector embeddings.
   The latter are compatible with a wide range of machine-learning algorithms and have proven versatile in disparate domains~\cite{Mikolov2013,Le2014}, and model classes, e.g.:
   \begin{itemize*}[label=]
   	\item Node2vec~\cite{Grover2016};
   	\item graph neural networks~\cite{velickovic2018,gilmer2017};
   	\item Isomap~\cite{Tenenbaum2000};
   	\item M-NMF~\cite{wang2017}.
   \end{itemize*}
   Among these, we distinguish graph neural networks (GNNs) \cite{velickovic2018,gilmer2017,morris2024} due to their wide-spread adoption, and ability to directly learn embeddings from graph-structured data.
   GNNs are predominantly trained in a supervised or semi-\-supervised fashion~\cite{kipf2017}; optimizing a node classification loss, potentially augmented by auxiliary terms, e.g. reconstruction error or noise contrastive estimate~\cite{velickovic2018a,chamberlain2017,chen2018c}.
   The resulting algorithms have proven to be efficient, and empirically successful when applied to tasks such as graph regression, node classification, and link prediction~\cite{dwivedi2023}.
   
   Comparatively, the use of GNNs for community detection has been relatively underexplored~\cite{su2024},
   despite the pivotal role that communities play in understanding the structure of biology, social, and economic networks~\cite{yang2012}.
   In this context, the applicability of GNNs is curtailed by their reliance on labeled data, which is often unavailable or requires cost-intensive curation.
   
   Classically, clustering algorithms have been used to address the community detection problem, e.g.: Girvan-Newman~\cite{girvan2002community}, Louvain~\cite{blondel2008fast}, and spectral clustering~\cite{pothen1990} algorithms.
   More recent approaches, e.g.~\cite{su2024,ji2017}, use GNN's to construct an embedding of the data, which is then post-processed using an off-\-the-\-shelf clustering algorithm, such as mean shift~\cite{Koohpayegani2021}, DBSCAN~\cite{Ester1996}, HDBSCAN~\cite{Malzer2020}, Birch~\cite{Zhang1996}, OPTICS~\cite{Ankerst1999}, AffinityPropagation~\cite{Sun2018}, AgglomerativeClustering~\cite{aggclust}.\\
   Our proposed approach improves on GNN approaches by offering a lightweight neural network model, that can be optimized efficiently without label supervision.
   
   %
   Recent studies citing Agrawal's work underscore its relevance across diverse biological research areas. For example, research highlighting the pitfalls of extreme dimensionality reduction in single-cell genomics-- \cite{Chari2023} leverages Agrawal's insights to critique conventional visualization techniques, advocating for targeted embedding strategies for more meaningful biological analysis. Similarly, an integrated single-cell dataset study of the hypothalamic paraventricular nucleus (PVN) \cite{Berkhout2024} applies Agrawal's principles to achieve a nuanced molecular and functional classification, revealing the complexity of neuroendocrine regulation. Additionally, advancements in brain-wide neuronal activity recording through blazed oblique plane microscopy, as detailed in a study from Nature \cite{Hoffmann2023}, demonstrate the utility of Agrawal's work in bridging cellular and macroscale understanding. This particular study challenges the prevailing view that higher resolution is invariably superior, illustrating the impact of Agrawal's contributions to enhancing our understanding of complex biological systems.

\section{Problem statement}

   We consider a graph given by a set of vertices, \(V = \{v_1, \dots, v_n\}\), where \(|V|=n\) is the order of the graph.
   Given a fixed, we represent the shortest path length between pairs of vertices as the matrix \(D \in \real_+^{n\times n}\).
   These shortest distances may be based on unitary distances between adjacent vertices but also on any positive distances between adjacent vertices.
   In the present study, we focus exclusively on undirected graphs; therefore, $D$ is symmetric. In the case of disconnected graphs, we assume the topological distance between vertices in different subgraphs is equal to the maximum distance between vertices in the same subgraph plus one.

   The problem is to assign an embedding, \(e_i\in\real^m\), to each vertex \(v_i\in V\), in such a way that the approximate equality
   \begin{equation} \label{main:requirement}
     d(e_i,e_j) \cong D_{i,j}
   \end{equation}
   holds for each \(i,j\in\{1, \dots, n\}\), where \(d(\cdot,\cdot)\) is a certain distance in \(\mathbb{R}^m\) , e.g., the Euclidean distance.

  \section{Method}

   \subsection{Embeddings as solution to optimization problem}

   Following \cite{agrawal2021}, we obtain the embeddings \(e_i\) \eqref{main:requirement} as a~solution to an optimization problem which minimizes the average discrepancy between \(d(e_i,e_j)\) and \(D_{i,j}\):
   \begin{equation} \label{opt:0}
     \left[ e_1, \dots, e_n \right] = \argmin_{[e_1, \dots, e_n]} \frac2{n(n-1)} \sum_{i=1}^n\sum_{j=1}^{i-1}
     L(d(e_i,e_j), D_{i,j}).
   \end{equation}
   These discrepancies, denoted above by \(L(\cdot,\cdot)\), may be absolute
   \begin{equation} \label{loss:abs}
     L(d(e_i,e_j), D_{i,j}) = (d(e_i,e_j)-D_{i,j})^2
   \end{equation}
   or relative
   \begin{equation} \label{loss:rel}
     L(d(e_i,e_j), D_{i,j}) = (d(e_i,e_j)/D_{i,j}-1)^2.
   \end{equation}

   \subsection{Regularized embeddings}

   We notice that we can identify a vertex by a vector of distances to all vertices in the graph, i.e., a row or column in the \(D\) matrix.
   This is because such a vector contains only one \(0\), and its position identifies the vertex of the question.
   Moreover, if two columns in the \(D\) matrix are similar, their respective vertices are in similar distance to other vertices and are usually close\comment{to one another}.
   Therefore, their respective embeddings should also be close.
   Therefore, we consider the embeddings as results of continuous transformations of the columns in the \(D\) matrix. This continuity is a form of regularization that prevents the optimization process \eqref{opt:0} from getting stuck in a~poor local minimum.

   We consider the above transformation in the form of the deep neural network,
   \begin{equation} \label{e=f}
     e_j = f(D_{\cdot,j}; \theta),
   \end{equation}
   where \(f\) denotes the network and \(\theta\) represents its weights.
   Then, to compute the embeddings, we need to optimize the weights \(\theta\) of the network:
   \begin{equation} \label{opt:1}
     \theta = \argmin_{\theta} \frac2{n(n-1)} \sum_{i=1}^n\sum_{j=1}^{i-1}
     L(d(f(D_{\cdot,i};\theta),f(D_{\cdot,j};\theta)), D_{i,j}).
   \end{equation}

   \subsection{Distance functions}

   \begin{figure}
     \centering
     \includegraphics[width=2cm]{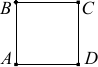}
     \caption{Simple square graph}
     \label{fig:square:graph}
   \end{figure}

   Let us consider the graph in Fig.~\ref{fig:square:graph} to motivate the need for flexible distance functions in the embedding space.
   The topological distance for adjacent vertices equals \(1\), and for diagonally-opposite vertices equals \(2\).
   Under an \(\ell_{2}\) distance, there is no possible arrangement of the embeddings \(e_{i}\) in $\real^m$ that preserves \(D_{i,j}\).
   If the distance between the adjacent vertex embeddings equals \(1\), the distances between both pairs of opposite vertex embeddings cannot equal \(2\)  {\it at the same time}.

   Let us consider a generalized distance formula
   \begin{equation} \label{dist:Euclid:kap}
     d(e_i, e_j) = \|e_i - e_j\|^\kappa,
   \end{equation}
   where \(\kappa \in [0, 2]\). For $\kappa=1$, \eqref{dist:Euclid:kap} represents the Euclidean distance.
   Treating $\kappa$ as a parameter, we optimize it jointly, along with the vertex embeddings of each graph.

   Coming back to the graph in Fig.~\ref{fig:square:graph}, let us consider $m=2$ and the most natural embeddings for the vertices: $e_A=[0, 0]^T$, $e_B=[0, 1]^T$, $e_C=[1, 1]^T$, $e_D=[1,0]^T$. It is seen that for $\kappa=2$, we have $d(e_i,e_j)=1$ for each pair of adjacent vertices and $d(e_i,e_j)=2$ for each pair of the opposite vertices. This exactly reflects the topological distances.

   For an arbitrary graph, variable $\kappa$ may improve the fitness of embeddings measured by \eqref{loss:abs} and by \eqref{loss:rel}.
   Additionally, optimizing $\kappa$ to minimize the discrepancy allows the model to adapt to the specific structure of the graph, which may be beneficial for the quality of the embeddings.

  \section{Experiments}

   Our experiments are centered on three key areas, which we have identified as particularly compelling during the development of our method.
   These encompass evaluating the quality of graph embeddings determined by the loss function, generating graphs with embeddings produced by our method, and identifying communities within graphs using our approach.
   The challenge of community detection can be assessed through the \textit{modularity} of the detected communities and by comparing them to ground truth communities.
   Since we did not restrict our benchmarks to sets of graphs exclusively comprised of community structures, most of our measurements are based on the \textit{modularity} metric.

   \subsection{Datasets}

   For\comment{ the purpose of} graph analysis and modularity measurement, we employed \textit{TUDataset} \cite{Morris+2020}, specifically opting for one graph set from each thematic category within the dataset.
   The selected graph sets are as follows: \textit{MUTAG (small molecules)}, \textit{ENZYMES (Bioinformatics)}, \textit{Cuneiform (Computer vision)}, \textit{IMDB-BINARY (Social networks)}, and \textit{SYNTHETIC (Synthetic)}.
   These sets comprise graphs of relatively small size and similar characteristics.

   Additionally, we utilized the \textit{CORA} dataset \cite{Collective+2008}, which features a single, large graph.
   For evaluating the community detection task, we employed two specific graphs representing real-world communities: the \textit{Zachary Karate Club} \cite{konect:ucidata-zachary} and \textit{American Football} \cite{girvan2002community}.
   To visualize how different parameters affect final embeddings, we employed \textit{American Revolution} graph \cite{konect:2017:brunson_revolution}.

   \subsection{Graph analysis}

   This section is dedicated to presenting the results regarding the performance of each analyzed method in preserving the actual topological distances between graph nodes within the embedding space.
   We introduce two metrics, RMSE and RMRSE (depending on the utilized loss function for generating embeddings), defined for a single graph as follows:

   \begin{align}
     RMSE  & = \sqrt{ \frac2{n(n-1)} \sum_{i=1}^n\sum_{j=1}^{i-1} ( d(e_i, e_j) - D_{i,j})^2 } \label{def:RMSE} \\
     RMRSE & = \sqrt{ \frac2{n(n-1)} \sum_{i=1}^n\sum_{j=1}^{i-1} ( d(e_i, e_j)/D_{i,j} - 1)^2 }.
     \label{def:RMRSE}
   \end{align}

   To analyze both the RMSE and RMRSE measures, we compared two approaches: a predefined \(\kappa=1\) with a value of 1.0 (Table \ref{table:Mean_loss_kappa_1.0}) and an automatically learned \(\kappa\) (Table \ref{table:Mean_loss_kappa_auto}).
   From both tables, we observe that the loss generally decreases as the dimensionality of embeddings increases.
   This observation is expected, as embeddings in higher-dimensional vector spaces can compress a graph's nodes to preserve topological distances for more complex graphs. However, for each dataset, there is a specific limit embedding dimension above which its further increase does not lead to any loss gain.


   As \comment{it is} seen in Table~\ref{table:Mean_loss_kappa_1.0}, the embeddings optimized directly provide us with worse results than those produced by the fitted neural network.
     This is especially visible for small embedding dimensions.
     The regularization introduced by the neural network prevents the optimization of the embeddings from getting stuck in local minima.
     For larger dimensions, the neural network does not introduce any improvement.
     The only exception is the CORE graph, for which the network could not learn a well-fitting transformation.

   Comparing Table \ref{table:Mean_loss_kappa_1.0} with Table \ref{table:Mean_loss_kappa_auto}, we note that the automatically learned \(\kappa\) parameter yields better results, usually by a large margin.

   In the Table \ref{table:auto_kappa} we present values of the automatically learned \(\kappa\) parameter.
    We see the \(\kappa\) usually increases with \(m\).
     This relation reflects the increasing difficulty in preserving the topological distances of vertices in embedding space of a decreasing dimension.
     Notably, for large \(m\), the optimal \(\kappa\) is usually well above \(1\).

   \begin{table}
     \centering
     \begin{adjustbox}{max width=\textwidth,keepaspectratio}
       \begin{tabular}{|c|c|c|c|c|c|c|c|c|}
         \toprule
         Method & Dataset & m=2 & m=3 & m=5 & m=10 & m=15 & m=30 & m=50                                                                                             \\
         \midrule
         \multirow{5}{*}{\rotatebox[origin=c]{90}{\tiny{RMSE Direct}}} & MUTAG & 0.38 ± 0.23 & 0.24 ± 0.04 & 0.24 ± 0.03 & 0.24 ± 0.03 & 0.24 ± 0.03 & 0.24 ± 0.03 & 0.24 ± 0.03  \\
          & Cuneiform & 0.58 ± 0.16 & 0.33 ± 0.13 & 0.17 ± 0.07 & 0.11 ± 0.02 & 0.11 ± 0.02 & 0.11 ± 0.02 & 0.11 ± 0.02                                                           \\
          & SYNTHETIC & 1.32 ± 0.04 & 0.94 ± 0.03 & 0.60 ± 0.01 & 0.44 ± 0.00 & 0.43 ± 0.00 & 0.43 ± 0.00 & 0.43 ± 0.00                                                           \\
          & ENZYMES & 0.53 ± 0.71 & 0.30 ± 0.44 & 0.22 ± 0.28 & 0.20 ± 0.15 & 0.20 ± 0.11 & 0.20 ± 0.10 & 0.19 ± 0.10                                                             \\
          & IMDB-BINARY & 0.38 ± 0.09 & 0.25 ± 0.06 & 0.15 ± 0.04 & 0.09 ± 0.03 & 0.07 ± 0.03 & 0.07 ± 0.04 & 0.07 ± 0.04                                                         \\
          & CORA & 3.30 & 2.47 & 1.80 & 1.17 & 0.90 & 0.57 & 0.45                                                                                                                 \\
         \midrule
         \multirow{5}{*}{\rotatebox[origin=c]{90}{\tiny{RMSE Neural}}} & MUTAG & 0.30 ± 0.06 & 0.26 ± 0.04 & 0.25 ± 0.04 & 0.25 ± 0.04 & 0.25 ± 0.03 & 0.25 ± 0.04 & 0.26 ± 0.04  \\
          & Cuneiform & 0.55 ± 0.14 & 0.33 ± 0.12 & 0.17 ± 0.07 & 0.12 ± 0.02 & 0.11 ± 0.02 & 0.11 ± 0.02 & 0.11 ± 0.02                                                           \\
          & SYNTHETIC & 1.20 ± 0.01 & 0.89 ± 0.01 & 0.60 ± 0.01 & 0.45 ± 0.00 & 0.44 ± 0.00 & 0.44 ± 0.00 & 0.44 ± 0.00                                                           \\
          & ENZYMES & 0.36 ± 0.22 & 0.28 ± 0.15 & 0.25 ± 0.13 & 0.24 ± 0.10 & 0.25 ± 0.18 & 0.26 ± 0.11 & 0.25 ± 0.11                                                             \\
          & IMDB-BINARY & 0.38 ± 0.09 & 0.26 ± 0.06 & 0.16 ± 0.04 & 0.09 ± 0.03 & 0.08 ± 0.03 & 0.07 ± 0.03 & 0.07 ± 0.03                                                         \\
          & CORA & 2.69 & 2.20 & 1.98 & 2.16 & 3.75 & 1.47 & 2.50                                                                                                                 \\
         \midrule
         \multirow{5}{*}{\rotatebox[origin=c]{90}{\tiny{RMRSE Direct}}} & MUTAG & 0.12 ± 0.03 & 0.10 ± 0.01 & 0.10 ± 0.01 & 0.10 ± 0.01 & 0.10 ± 0.01 & 0.10 ± 0.01 & 0.10 ± 0.01 \\
          & Cuneiform & 0.24 ± 0.04 & 0.14 ± 0.05 & 0.09 ± 0.03 & 0.07 ± 0.01 & 0.07 ± 0.01 & 0.07 ± 0.01 & 0.07 ± 0.01                                                           \\
          & SYNTHETIC & 0.38 ± 0.01 & 0.29 ± 0.01 & 0.21 ± 0.00 & 0.17 ± 0.00 & 0.17 ± 0.00 & 0.17 ± 0.00 & 0.17 ± 0.00                                                           \\
          & ENZYMES & 0.15 ± 0.05 & 0.10 ± 0.04 & 0.09 ± 0.03 & 0.09 ± 0.03 & 0.09 ± 0.03 & 0.09 ± 0.03 & 0.09 ± 0.03                                                             \\
          & IMDB-BINARY & 0.29 ± 0.04 & 0.20 ± 0.04 & 0.12 ± 0.03 & 0.08 ± 0.02 & 0.06 ± 0.03 & 0.06 ± 0.03 & 0.06 ± 0.03                                                         \\
          & CORA & 0.44 & 0.34 & 0.23 & 0.12 & 0.09 & 0.08 & 0.08                                                                                                                 \\
         \midrule
         \multirow{5}{*}{\rotatebox[origin=c]{90}{\tiny{RMRSE Neural}}} & MUTAG & 0.11 ± 0.02 & 0.10 ± 0.01 & 0.10 ± 0.01 & 0.10 ± 0.01 & 0.10 ± 0.01 & 0.10 ± 0.01 & 0.10 ± 0.01 \\
          & Cuneiform & 0.24 ± 0.04 & 0.14 ± 0.05 & 0.09 ± 0.03 & 0.07 ± 0.01 & 0.07 ± 0.01 & 0.07 ± 0.01 & 0.07 ± 0.01                                                           \\
          & SYNTHETIC & 0.36 ± 0.00 & 0.28 ± 0.00 & 0.21 ± 0.00 & 0.18 ± 0.00 & 0.18 ± 0.00 & 0.18 ± 0.00 & 0.18 ± 0.00                                                           \\
          & ENZYMES & 0.14 ± 0.04 & 0.10 ± 0.04 & 0.09 ± 0.03 & 0.09 ± 0.03 & 0.09 ± 0.03 & 0.09 ± 0.03 & 0.09 ± 0.03                                                             \\
          & IMDB-BINARY & 0.29 ± 0.04 & 0.20 ± 0.04 & 0.12 ± 0.03 & 0.08 ± 0.02 & 0.07 ± 0.03 & 0.06 ± 0.03 & 0.06 ± 0.03                                                         \\
          & CORA & 0.37 & 0.29 & 0.23 & 0.21 & 0.21 & 0.19 & 0.18                                                                                                                 \\
         \bottomrule
       \end{tabular}
     \end{adjustbox}
     \caption{Mean loss function results for \(\kappa=1.0\). Methods: RMSE=absolute loss, RMRSE=relative loss, Direct=embeddings optimized directly, Neural=embeddings regularized with a neural network}
     \label{table:Mean_loss_kappa_1.0}
   \end{table}

   \begin{table}
     \centering
     \begin{adjustbox}{max width=\textwidth,keepaspectratio}
       \begin{tabular}{|c|c|c|c|c|c|c|c|c|}
         \toprule
         Method & Dataset & m=2 & m=3 & m=5 & m=10 & m=15 & m=30 & m=50                                                                                             \\
         \midrule
         \multirow{5}{*}{\rotatebox[origin=c]{90}{\tiny{RMSE Direct}}} & MUTAG & 0.35 ± 0.19 & 0.21 ± 0.04 & 0.15 ± 0.03 & 0.04 ± 0.03 & 0.02 ± 0.01 & 0.02 ± 0.01 & 0.02 ± 0.00  \\
          & Cuneiform & 0.50 ± 0.13 & 0.30 ± 0.11 & 0.16 ± 0.08 & 0.05 ± 0.02 & 0.03 ± 0.02 & 0.01 ± 0.00 & 0.01 ± 0.00                                                           \\
          & SYNTHETIC & 1.13 ± 0.04 & 0.87 ± 0.03 & 0.60 ± 0.01 & 0.39 ± 0.00 & 0.29 ± 0.00 & 0.24 ± 0.00 & 0.24 ± 0.00                                                           \\
          & ENZYMES & 0.39 ± 0.33 & 0.25 ± 0.18 & 0.17 ± 0.10 & 0.10 ± 0.07 & 0.08 ± 0.05 & 0.06 ± 0.04 & 0.06 ± 0.04                                                             \\
          & IMDB-BINARY & 0.23 ± 0.10 & 0.17 ± 0.07 & 0.12 ± 0.05 & 0.06 ± 0.04 & 0.03 ± 0.03 & 0.01 ± 0.02 & 0.02 ± 0.02                                                         \\
          & CORA & 3.25 & 2.56 & 1.76 & 0.93 & 0.69 & 0.48 & 0.39                                                                                                                 \\
         \midrule
         \multirow{5}{*}{\rotatebox[origin=c]{90}{\tiny{RMSE Neural}}} & MUTAG & 0.30 ± 0.06 & 0.24 ± 0.04 & 0.20 ± 0.04 & 0.19 ± 0.03 & 0.20 ± 0.05 & 0.20 ± 0.04 & 0.20 ± 0.07  \\
          & Cuneiform & 0.43 ± 0.09 & 0.29 ± 0.09 & 0.16 ± 0.07 & 0.09 ± 0.03 & 0.08 ± 0.02 & 0.08 ± 0.03 & 0.08 ± 0.04                                                           \\
          & SYNTHETIC & 0.84 ± 0.03 & 0.73 ± 0.01 & 0.59 ± 0.01 & 0.40 ± 0.01 & 0.35 ± 0.02 & 0.34 ± 0.02 & 0.34 ± 0.02                                                           \\
          & ENZYMES & 0.38 ± 0.20 & 0.31 ± 0.16 & 0.26 ± 0.13 & 0.26 ± 0.17 & 0.26 ± 0.19 & 0.26 ± 0.16 & 0.28 ± 0.21                                                             \\
          & IMDB-BINARY & 0.21 ± 0.06 & 0.18 ± 0.05 & 0.13 ± 0.04 & 0.08 ± 0.03 & 0.06 ± 0.03 & 0.06 ± 0.03 & 0.06 ± 0.03                                                         \\
          & CORA & 2.29 & 2.84 & 2.40 & 1.84 & 1.75 & 2.45 & 1.25                                                                                                                 \\
         \midrule
         \multirow{5}{*}{\rotatebox[origin=c]{90}{\tiny{RMRSE Direct}}} & MUTAG & 0.10 ± 0.03 & 0.07 ± 0.01 & 0.05 ± 0.01 & 0.01 ± 0.01 & 0.00 ± 0.00 & 0.01 ± 0.00 & 0.01 ± 0.00 \\
          & Cuneiform & 0.22 ± 0.03 & 0.14 ± 0.05 & 0.07 ± 0.04 & 0.02 ± 0.01 & 0.01 ± 0.01 & 0.00 ± 0.00 & 0.00 ± 0.00                                                           \\
          & SYNTHETIC & 0.36 ± 0.01 & 0.28 ± 0.01 & 0.21 ± 0.00 & 0.12 ± 0.00 & 0.09 ± 0.00 & 0.08 ± 0.00 & 0.08 ± 0.00                                                           \\
          & ENZYMES & 0.14 ± 0.05 & 0.09 ± 0.04 & 0.06 ± 0.03 & 0.04 ± 0.02 & 0.03 ± 0.02 & 0.03 ± 0.02 & 0.03 ± 0.02                                                             \\
          & IMDB-BINARY & 0.16 ± 0.06 & 0.14 ± 0.05 & 0.10 ± 0.04 & 0.04 ± 0.03 & 0.02 ± 0.02 & 0.01 ± 0.01 & 0.01 ± 0.02                                                         \\
          & CORA & 0.42 & 0.29 & 0.19 & 0.11 & 0.09 & 0.07 & 0.06                                                                                                                 \\
         \midrule
         \multirow{5}{*}{\rotatebox[origin=c]{90}{\tiny{RMRSE Neural}}} & MUTAG & 0.10 ± 0.03 & 0.07 ± 0.01 & 0.05 ± 0.01 & 0.01 ± 0.01 & 0.00 ± 0.00 & 0.01 ± 0.00 & 0.01 ± 0.00 \\
          & Cuneiform & 0.22 ± 0.03 & 0.14 ± 0.05 & 0.07 ± 0.04 & 0.02 ± 0.01 & 0.01 ± 0.01 & 0.00 ± 0.00 & 0.00 ± 0.00                                                           \\
          & SYNTHETIC & 0.36 ± 0.01 & 0.28 ± 0.01 & 0.21 ± 0.00 & 0.12 ± 0.00 & 0.09 ± 0.00 & 0.08 ± 0.00 & 0.08 ± 0.00                                                           \\
          & ENZYMES & 0.14 ± 0.05 & 0.09 ± 0.04 & 0.06 ± 0.03 & 0.04 ± 0.02 & 0.03 ± 0.02 & 0.03 ± 0.02 & 0.03 ± 0.02                                                             \\
          & IMDB-BINARY & 0.16 ± 0.06 & 0.14 ± 0.05 & 0.10 ± 0.04 & 0.04 ± 0.03 & 0.02 ± 0.02 & 0.01 ± 0.01 & 0.01 ± 0.02                                                         \\
          & CORA & 0.42 & 0.29 & 0.19 & 0.11 & 0.09 & 0.07 & 0.06                                                                                                                 \\
         \bottomrule
       \end{tabular}
     \end{adjustbox}
     \caption{Mean loss function results for \(\kappa\) optimized along with embeddings. Method: see Table~\ref{table:Mean_loss_kappa_1.0}}
     \label{table:Mean_loss_kappa_auto}
   \end{table}

   \begin{table}
     \centering
     \begin{adjustbox}{max width=\textwidth,keepaspectratio}
       \begin{tabular}{|c|c|c|c|c|c|c|c|c|}
         \toprule
         Method & Dataset & m=2 & m=3 & m=5 & m=10 & m=15 & m=30 & m=50                                                                                             \\
         \midrule
         \multirow{5}{*}{\rotatebox[origin=c]{90}{\tiny{RMSE Direct}}} & MUTAG & 1.02 ± 0.06 & 1.13 ± 0.04 & 1.33 ± 0.09 & 1.81 ± 0.16 & 1.89 ± 0.07 & 1.88 ± 0.07 & 1.85 ± 0.08  \\
          & Cuneiform & 0.62 ± 0.11 & 0.81 ± 0.14 & 1.06 ± 0.11 & 1.25 ± 0.02 & 1.34 ± 0.04 & 1.45 ± 0.11 & 1.46 ± 0.11                                                           \\
          & SYNTHETIC & 0.70 ± 0.01 & 0.72 ± 0.01 & 0.84 ± 0.01 & 1.41 ± 0.00 & 1.81 ± 0.01 & 1.95 ± 0.02 & 1.95 ± 0.02                                                           \\
          & ENZYMES & 1.02 ± 0.12 & 1.08 ± 0.07 & 1.19 ± 0.11 & 1.47 ± 0.22 & 1.66 ± 0.22 & 1.74 ± 0.19 & 1.70 ± 0.19                                                             \\
          & IMDB-BINARY & 0.39 ± 0.16 & 0.54 ± 0.24 & 0.75 ± 0.32 & 1.06 ± 0.45 & 1.32 ± 0.41 & 1.50 ± 0.29 & 1.48 ± 0.24                                                         \\
          & CORA & 1.24 & 1.22 & 1.20 & 1.18 & 1.16 & 1.24 & 1.27                                                                                                                 \\
         \midrule
         \multirow{5}{*}{\rotatebox[origin=c]{90}{\tiny{RMSE Neural}}} & MUTAG & 1.05 ± 0.03 & 1.15 ± 0.06 & 1.24 ± 0.08 & 1.27 ± 0.09 & 1.24 ± 0.09 & 1.24 ± 0.09 & 1.24 ± 0.10  \\
          & Cuneiform & 0.62 ± 0.13 & 0.83 ± 0.13 & 1.06 ± 0.10 & 1.20 ± 0.03 & 1.22 ± 0.04 & 1.23 ± 0.05 & 1.23 ± 0.06                                                           \\
          & SYNTHETIC & 0.50 ± 0.05 & 0.64 ± 0.05 & 0.90 ± 0.03 & 1.30 ± 0.07 & 1.42 ± 0.09 & 1.44 ± 0.09 & 1.45 ± 0.07                                                           \\
          & ENZYMES & 1.03 ± 0.13 & 1.11 ± 0.11 & 1.19 ± 0.12 & 1.20 ± 0.13 & 1.20 ± 0.13 & 1.21 ± 0.13 & 1.20 ± 0.13                                                             \\
          & IMDB-BINARY & 0.45 ± 0.14 & 0.59 ± 0.19 & 0.79 ± 0.25 & 1.04 ± 0.31 & 1.14 ± 0.27 & 1.20 ± 0.23 & 1.20 ± 0.22                                                         \\
          & CORA & 0.83 & 1.38 & 1.36 & 1.15 & 1.22 & 0.92 & 0.91                                                                                                                 \\
         \midrule
         \multirow{5}{*}{\rotatebox[origin=c]{90}{\tiny{RMRSE Direct}}} & MUTAG & 1.10 ± 0.03 & 1.19 ± 0.03 & 1.40 ± 0.07 & 1.87 ± 0.12 & 1.94 ± 0.06 & 1.91 ± 0.05 & 1.84 ± 0.05 \\
          & Cuneiform & 0.76 ± 0.08 & 0.97 ± 0.08 & 1.13 ± 0.06 & 1.26 ± 0.03 & 1.35 ± 0.05 & 1.45 ± 0.11 & 1.46 ± 0.11                                                           \\
          & SYNTHETIC & 0.69 ± 0.01 & 0.81 ± 0.01 & 1.11 ± 0.00 & 1.59 ± 0.00 & 1.90 ± 0.01 & 1.96 ± 0.00 & 1.91 ± 0.01                                                           \\
          & ENZYMES & 1.00 ± 0.08 & 1.09 ± 0.05 & 1.26 ± 0.11 & 1.63 ± 0.20 & 1.80 ± 0.18 & 1.83 ± 0.15 & 1.76 ± 0.13                                                             \\
          & IMDB-BINARY & 0.39 ± 0.16 & 0.56 ± 0.24 & 0.82 ± 0.35 & 1.15 ± 0.48 & 1.37 ± 0.43 & 1.52 ± 0.29 & 1.49 ± 0.25                                                         \\
          & CORA & 1.01 & 1.06 & 1.11 & 1.11 & 1.14 & 1.31 & 1.50                                                                                                                 \\
         \midrule
         \multirow{5}{*}{\rotatebox[origin=c]{90}{\tiny{RMRSE Neural}}} & MUTAG & 1.11 ± 0.02 & 1.19 ± 0.03 & 1.30 ± 0.05 & 1.34 ± 0.06 & 1.35 ± 0.06 & 1.33 ± 0.06 & 1.33 ± 0.07 \\
          & Cuneiform & 0.76 ± 0.08 & 0.97 ± 0.08 & 1.13 ± 0.06 & 1.23 ± 0.03 & 1.24 ± 0.04 & 1.24 ± 0.04 & 1.25 ± 0.04                                                           \\
          & SYNTHETIC & 0.66 ± 0.04 & 0.85 ± 0.02 & 1.12 ± 0.01 & 1.50 ± 0.05 & 1.59 ± 0.07 & 1.62 ± 0.07 & 1.63 ± 0.04                                                           \\
          & ENZYMES & 0.98 ± 0.08 & 1.09 ± 0.07 & 1.21 ± 0.10 & 1.27 ± 0.12 & 1.27 ± 0.12 & 1.27 ± 0.11 & 1.27 ± 0.11                                                             \\
          & IMDB-BINARY & 0.43 ± 0.12 & 0.60 ± 0.19 & 0.85 ± 0.28 & 1.14 ± 0.35 & 1.24 ± 0.32 & 1.29 ± 0.27 & 1.30 ± 0.26                                                         \\
          & CORA & 1.13 & 1.09 & 0.91 & 1.34 & 1.33 & 1.00 & 1.11                                                                                                                 \\
         \bottomrule
       \end{tabular}
     \end{adjustbox}
     \caption{Mean learned \(\kappa\)}
     \label{table:auto_kappa}
   \end{table}

   \subsection{Graph drawing}

\begin{figure}
     \hspace{-3em}
     \includegraphics[width=14cm]{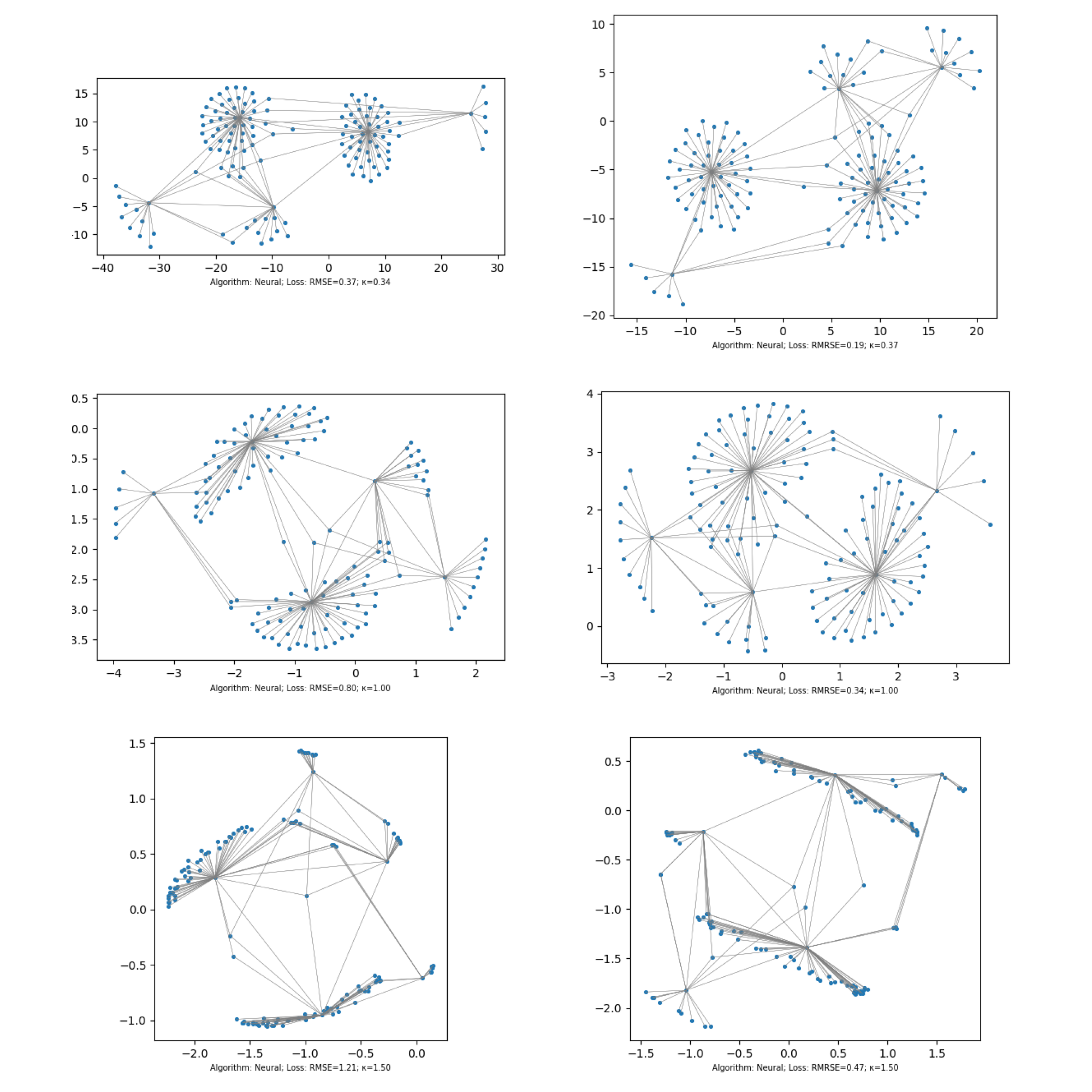}
     \caption{American Revolution \cite{konect:2017:brunson_revolution} graph visual representation ($m=2$). {\it Left:} absolute error minimized. {\it Right:} relative error minimized. {\it Top:} $\kappa=\textit{auto}\cong0.4$, {\it Middle:} $\kappa=1$, {\it Bottom:} $\kappa=1.5$.}
     \label{fig:drawing}
   \end{figure}

   Figure \ref{fig:drawing} illustrates\comment{ graph} vertex embeddings for different $\kappa$ and loss types (absolute or relative). We selected the American Revolution \cite{konect:2017:brunson_revolution} graph for visualization purposes, as the graph perfectly shows local node communities being represented in the embeddings' space. It is seen that minimization of the relative loss puts more emphasis on the local graph structure, while minimization of the absolute loss focuses more on its global structure. Also, with higher $\kappa$ comes a tendency to compress local clusters of vertices.

   \subsection{Community detection in graphs}

   \subsubsection{Methodology.}

   Our study aimed to identify communities within graphs using a diverse array of unsupervised clustering algorithms applied to the embeddings of graph nodes generated through our methods.
   The clustering algorithms employed encompassed MeanShift \cite{Koohpayegani2021}, DBSCAN \cite{Ester1996}, HDBSCAN \cite{Malzer2020}, Birch \cite{Zhang1996}, OPTICS \cite{Ankerst1999}, AffinityPropagation \cite{Sun2018} and AgglomerativeClustering \cite{aggclust}

   We assessed the modularity score for all graphs, as this metric does not necessitate a graph with defined communities.
   For the Zachary Karate Club and American Football graphs, we additionally examined the ARS (adjusted rand index) and NMI (normalized mutual information) scores, which are valuable metrics for assessing how effectively an algorithm detects known communities.

   We compared the results of our method with five widely used community detection algorithms: greedy modularity communities \cite{clauset2004greedy}, Louvain communities \cite{blondel2008fast}, Kernighan Lin bisection \cite{kernighan1970efficient}, Girvan Newman \cite{girvan2002community}, and asynchronous label propagation algorithm (asyn LPA)\cite{Raghavan_2007}.

   \subsubsection{Results.}

   Table \ref{table:zachary} presents the Zachary Karate Club graph results.
   Our method outperforms other community detection algorithms in terms of ARS and NMI scores.
   Interestingly, the modularity score in this case does not align with the performance of the best method.

   Table \ref{table:football} exhibits similar results but for the American Football graph.
   Our method also performs admirably, although slightly worse than the Girvan Newman method.

   The modularity results for the TUDataset are detailed in Table \ref{table:mean_modularity_emb}.
   We compared these results with\comment{ those of} other algorithms, as shown in Table \ref{table:networkx_algs}.
   Our method achieves competitive modularity scores, but not the highest ones.
   Notably, the method of designating the embedding does not impact the achieved modularity.

   \begin{table}
     \centering
     \begin{tabular}{|lrrrrrrr|}
       \toprule
       Embed. method & Error & \(\kappa\) & \(m\) & Clustering algorithm & ARS & NMI & Modularity \\
       \midrule
       - & - & - & - & Greedy Modularity Communities & 0.57 & 0.56 & 0.38                         \\
       - & - & - & - & Louvain communities & 0.51 & 0.60 & \textbf{0.42}                          \\
       - & - & - & - & Kernighan Lin Bisection & 0.77 & 0.68 & 0.37                               \\
       - & - & - & - & Girvan Newman & 0.77 & 0.73 & 0.36                                         \\
       - & - & - & - & Asyn Lpa Communities & 0.66 & 0.65 & 0.38                                  \\
       \midrule
       Direct RMSE & 0.45 & 0.72 & 2 & MeanShift & 0.88 & 0.84 & 0.37                             \\
       Direct RMRSE & 0.19 & 0.73 & 5 & MeanShift & 0.88 & 0.84 & 0.37                            \\
       Direct RMSE & 0.54 & 1.00 & 2 & MeanShift & 0.88 & 0.84 & 0.37                             \\
       Direct RMRSE & 0.26 & 1.00 & 2 & AffinityPropagation & 0.88 & 0.84 & 0.37                  \\
       Direct RMSE & 0.76 & 1.50 & 2 & MeanShift & 0.88 & 0.84 & 0.37                             \\
       Direct RMRSE & 0.37 & 1.50 & 2 & MeanShift & 0.88 & 0.84 & 0.37                            \\
       \midrule
       Neural RMSE & 0.38 & 0.72 & 3 & MeanShift & 0.88 & 0.84 & 0.37                             \\
       Neural RMRSE & 0.19 & 0.74 & 10 & MeanShift & 0.88 & 0.84 & 0.37                           \\
       Neural RMSE & 0.52 & 1.00 & 2 & MeanShift & 0.88 & 0.84 & 0.37                             \\
       Neural RMRSE & 0.16 & 1.00 & 5 & MeanShift & 0.88 & 0.84 & 0.37                            \\
       Neural RMSE & 0.74 & 1.50 & 2 & MeanShift & 0.88 & 0.84 & 0.37                             \\
       Neural RMRSE & 0.36 & 1.50 & 2 & MeanShift & \textbf{1.00} & \textbf{1.00} & 0.36          \\
       \bottomrule
     \end{tabular}
     \caption{Community detection in Zachary Karate Club}
     \label{table:zachary}
   \end{table}

   \begin{table}
     \centering
     \begin{tabular}{|lrrrrrrr|}
       \toprule
       Embed. method & Error & \(\kappa\) & \(m\) & Clustering algorithm & ARS & NMI & Modularity \\
       \midrule
       - & - & - & - & Greedy Modularity Communities & 0.47 & 0.70 & 0.55                         \\
       - & - & - & - & Louvain Communities & 0.81 & 0.89 & \textbf{0.60}                          \\
       - & - & - & - & Kernighan Lin Bisection & 0.14 & 0.38 & 0.35                               \\
       - & - & - & - & Girvan Newman & \textbf{0.92} & \textbf{0.94} & 0.36                       \\
       - & - & - & - & Asyn Lpa Communities & 0.75 & 0.87 & 0.58                                  \\
       \midrule
       Direct RMSE & 0.46 & 0.54 & 10 & AffinityPropagation & 0.78 & 0.86 & 0.58                  \\
       Direct RMRSE & 0.24 & 0.51 & 5 & HDBSCAN & 0.89 & 0.92 & 0.58                              \\
       Direct RMSE & 0.36 & 1.00 & 15 & AffinityPropagation & 0.86 & 0.92 & 0.58                  \\
       Direct RMRSE & 0.18 & 1.00 & 15 & AffinityPropagation & 0.86 & 0.92 & 0.58                 \\
       Direct RMSE & 0.51 & 1.50 & 5 & MeanShift & 0.83 & 0.89 & 0.51                             \\
       Direct RMRSE & 0.13 & 1.50 & 50 & AffinityPropagation & 0.78 & 0.87 & 0.54                 \\
       \midrule
       Neural RMSE & 0.46 & 0.55 & 10 & AffinityPropagation & 0.84 & 0.89 & 0.59                  \\
       Neural RMRSE & 0.24 & 0.52 & 10 & HDBSCAN & 0.81 & 0.89 & 0.52                             \\
       Neural RMSE & 0.36 & 1.00 & 15 & AffinityPropagation & 0.86 & 0.91 & 0.58                  \\
       Neural RMRSE & 0.18 & 1.00 & 30 & AffinityPropagation & 0.86 & 0.92 & 0.58                 \\
       Neural RMSE & 0.33 & 1.50 & 10 & AffinityPropagation & 0.81 & 0.90 & 0.55                  \\
       Neural RMRSE & 0.14 & 1.50 & 15 & AffinityPropagation & 0.89 & 0.89 & 0.57                 \\
       \bottomrule
     \end{tabular}
     \caption{Community detection in American Football}
     \label{table:football}
   \end{table}
   \begin{table}[t]
     \centering
     \begin{tabular}{|c|c|c|c|c|c|c|c|}
       \toprule
       Dataset & GMC & LC & GN & KLB & ALC                                                           \\
       \midrule
       MUTAG & {\bf 0.46 ± 0.06} & {\bf 0.46 ± 0.06} & {\bf 0.46 ± 0.06} & 0.34 ± 0.05 & 0.41 ± 0.05 \\
       Cuneiform & {\bf 0.53 ± 0.03} & {\bf 0.53 ± 0.03} & 0.47 ± 0.07 & 0.30 ± 0.07 & 0.47 ± 0.10   \\
       SYNTHETIC & {\bf 0.48 ± 0.00} & 0.47 ± 0.01 & 0.44 ± 0.00 & 0.31 ± 0.01 & 0.35 ± 0.03         \\
       ENZYMES & 0.57 ± 0.11 & {\bf 0.58 ± 0.12} & {\bf 0.58 ± 0.13} & 0.40 ± 0.08 & 0.54 ± 0.12     \\
       IMDB-BINARY & {\bf 0.30 ± 0.16} & {\bf 0.30 ± 0.16} & 0.27 ± 0.16 & 0.20 ± 0.12 & 0.25 ± 0.17 \\
       CORA & 0.81 & {\bf 0.82} & 0.81 & 0.41 & 0.5                                               \\
       \bottomrule
     \end{tabular}
     \caption{Mean modularity -- graph community detection algorithms}
     \label{table:networkx_algs}

     \centering
     \begin{adjustbox}{max width=\textwidth,keepaspectratio}
       \begin{tabular}{|c|c|c|c|c|c|c|c|c|}
         \toprule
          & \multicolumn{4}{|c|}{\(\kappa=1\)}                 & \multicolumn{4}{|c|}{\(\kappa\)=auto}                                                                   \\
         \midrule
         Method & Dataset & Clustering algorithm & m & Modularity & Dataset & Clustering algorithm & m & Modularity                                                  \\
         \midrule
         \multirow{5}{*}{\rotatebox[origin=c]{90}{\tiny{RMSE Direct}}} & MUTAG & AffinityPropagation & 50 & 0.43 ± 0.05 & MUTAG & AffinityPropagation & 5 & 0.44 ± 0.05  \\
          & Cuneiform & AffinityPropagation & 2 & 0.44 ± 0.12 & Cuneiform & AffinityPropagation & 10 & 0.52 ± 0.07                                                       \\
          & SYNTHETIC & AffinityPropagation & 5 & 0.39 ± 0.02 & SYNTHETIC & AffinityPropagation & 5 & 0.39 ± 0.01                                                        \\
          & ENZYMES & AffinityPropagation & 10 & 0.54 ± 0.13 & ENZYMES & AffinityPropagation & 2 & 0.54 ± 0.13                                                           \\
          & IMDB-BINARY & MeanShift & 2 & 0.26 ± 0.16 & IMDB-BINARY & HDBSCAN & 3 & 0.25 ± 0.17                                                                          \\
          & CORA & AffinityPropagation & 30 & 0.54 & CORA & AffinityPropagation & 50 & 0.57                                                                              \\
         \midrule
         \multirow{5}{*}{\rotatebox[origin=c]{90}{\tiny{RMSE Neural}}} & MUTAG & AffinityPropagation & 2 & 0.44 ± 0.05 & MUTAG & AffinityPropagation & 3 & 0.44 ± 0.05   \\
          & Cuneiform & AffinityPropagation & 2 & 0.52 ± 0.07 & Cuneiform & AffinityPropagation & 15 & 0.51 ± 0.06                                                       \\
          & SYNTHETIC & AffinityPropagation & 5 & 0.40 ± 0.01 & SYNTHETIC & AffinityPropagation & 5 & 0.40 ± 0.01                                                        \\
          & ENZYMES & AffinityPropagation & 2 & 0.55 ± 0.12 & ENZYMES & AffinityPropagation & 2 & 0.55 ± 0.12                                                            \\
          & IMDB-BINARY & MeanShift & 2 & 0.26 ± 0.16 & IMDB-BINARY & AffinityPropagation & 2 & 0.26 ± 0.18                                                              \\
          & CORA & AffinityPropagation & 30 & 0.51 & CORA & MeanShift & 3 & 0.60                                                                                         \\
         \midrule
         \multirow{5}{*}{\rotatebox[origin=c]{90}{\tiny{RMRSE Direct}}} & MUTAG & AffinityPropagation & 5 & 0.44 ± 0.05 & MUTAG & AffinityPropagation & 3 & 0.44 ± 0.05  \\
          & Cuneiform & AffinityPropagation & 10 & 0.53 ± 0.03 & Cuneiform & AffinityPropagation & 5 & 0.51 ± 0.09                                                       \\
          & SYNTHETIC & AffinityPropagation & 10 & 0.41 ± 0.00 & SYNTHETIC & AffinityPropagation & 5 & 0.40 ± 0.02                                                       \\
          & ENZYMES & AffinityPropagation & 5 & 0.55 ± 0.13 & ENZYMES & AffinityPropagation & 2 & 0.55 ± 0.12                                                            \\
          & IMDB-BINARY & HDBSCAN & 3 & 0.25 ± 0.17 & IMDB-BINARY & AffinityPropagation & 2 & 0.26 ± 0.18                                                                \\
          & CORA & AffinityPropagation & 30 & 0.64 & CORA & AffinityPropagation & 30 & 0.64                                                                              \\
         \midrule
         \multirow{5}{*}{\rotatebox[origin=c]{90}{\tiny{RMRSE Neural}}} & MUTAG & AffinityPropagation & 15 & 0.44 ± 0.05 & MUTAG & AffinityPropagation & 3 & 0.44 ± 0.05 \\
          & Cuneiform & AffinityPropagation & 15 & 0.53 ± 0.03 & Cuneiform & AffinityPropagation & 5 & 0.51 ± 0.08                                                       \\
          & SYNTHETIC & AffinityPropagation & 10 & 0.41 ± 0.01 & SYNTHETIC & AffinityPropagation & 5 & 0.40 ± 0.01                                                       \\
          & ENZYMES & AffinityPropagation & 2 & 0.55 ± 0.12 & ENZYMES & AffinityPropagation & 2 & 0.55 ± 0.12                                                            \\
          & IMDB-BINARY & HDBSCAN & 3 & 0.25 ± 0.17 & IMDB-BINARY & AffinityPropagation & 2 & 0.26 ± 0.18                                                                \\
          & CORA & AffinityPropagation & 50 & 0.58 & CORA & AffinityPropagation & 50 & 0.56                                                                              \\
         \bottomrule
       \end{tabular}
     \end{adjustbox}
     \caption{Best mean modularity in different datasets}
     \label{table:mean_modularity_emb}
   \end{table}

  \section{Conclusions}

     In this paper, we introduced a regularization method for graph vertex embeddings that preserves distances in the graph.
     This method uses a neural network to transform a column of the distance matrix into the embedding.
     In our experimental study, this regularization significantly improved the embeddings, especially when their dimension was low.

     We also introduced a generalized measure of distance between the embeddings.
     With our proposed measure, the error of distance preservation by the embeddings was reduced by a large margin.

  Finally, we performed a study on community detection in graphs, in which we compared results obtained by combining graph embeddings and clustering methods for numerical data with community detection algorithms dedicated to graphs. The analyzed combination achieved competitive results, although it yielded the best results only in the case of Zachary Karate Club.



\end{document}